\documentstyle[multicol,prb,aps,ifthen,epsfig]{revtex}

\newcommand{\wide}[2]{                                                        %
\end{multicols}                                                               %
\widetext                                                                     %
\noindent                                                                     %
\ifthenelse{\equal{#1}{t}}                                                    %
{}                                                                            %
{                                                                             %
\raisebox{0.1in}[0in][0.02in]{$\rule{3.575in}{0.002in}                        %
\rule{0.002in}{0.08in}$}                                                      %
}                                                                             %
#2                                                                            %
\ifthenelse{\equal{#1}{b}}                                                    %
{}                                                                            %
{                                                                             %
{\raisebox{-0.1in}[0in][0.02in]                                               %
{\hspace{3.575in}$\rule{0.002in}{0.08in}                                      %
\rule[0.08in]{3.575in}{0.002in}$}                                             %
}                                                                             %
}                                                                             %
\begin{multicols}{2}                                                          %
\noindent                                                                     %
}                                                                             %

\begin{document}

\title{Weak localization in ferromagnets with spin-orbit interaction}
\author{V.~K.~Dugaev$^{1,2}$\cite{email}, P.~Bruno$^1$, and J.~Barna\'s$^{3,4}$}
\address{$^1$Max-Planck-Institut f\"ur Mikrostrukturphysik,
Weinberg 2, D-06120 Halle, Germany\\
$^2$Institute of Materials Science Problems, Vilde~5, 58001 Chernovtsy, Ukraine\\
$^3$Department of Physics, A.~Mickiewicz University, ul.~Umultowska~85,
61-614~Pozna\'n, Poland \\
$^4$Institute of Molecular Physics, Polish Academy of Sciences,\\
ul.~M.~Smoluchowskiego~17, 60-179~Pozna\'n, Poland }
\date{\today }
\maketitle

\begin{abstract}

Weak localization corrections to conductivity of ferromagnetic systems
are studied theoretically in the case when spin-orbit interaction plays a 
significant role.
Two cases are analyzed in detail: (i) the case when the spin-orbit interaction 
is due to scattering from impurities,
and (ii) the case when the spin-orbit interaction
results from reduced dimensionality of the system
and is of the Bychkov-Rashba type. 
Results of the analysis show that the localization corrections  
to conductivity of ferromagnetic metals lead to a negative 
magnetoresistance -- also in the presence of the spin-orbit scattering.
Positive magnetoresistance
due to weak antilocalization, typical of nonmagnetic systems,
does not occur in ferromagnetic systems. 
In the case of two-dimensional ferromagnets, the quantum
corrections depend on the magnetization orientation 
with respect to the plane of the system. 

\end{abstract}

\begin{multicols}{2}

\section{Introduction}

Owing to the giant magnetoresistance effect 
discovered in artificially layered structures,\cite{baibich,binasch}
transport properties of low-dimensional magnetic systems were extensively
studied in the past decade.
The huge interest was stimulated by applications of the effect
in magnetoelectronics, particularly in read/write heads,
field sensors, random access memory elements, and others.\cite{prinz}
Since the effect exists also at room temperatures (which is important
for applications), there was only a little interest
in the low temperature regime, where
quantum corrections to conductivity
may play a certain role. 
This regime, however, may be important in view of possible applications
of metallic and/or semiconducting magnetic systems
in spintronics\cite{spintronics} and quantum computing.\cite{loss}

The quantum corrections to electrical conductivity, related to scattering of 
electrons from impurities in nonmagnetic metals and doped semiconductors,
were extensively studied in the past two
decades.\cite{alt82,lee85,alt85,bergmann}
However, the problem of quantum corrections in ferromagnetic metals is
still unexplored. Only a few works on this subject can be found in the
relevant literature. These include two 
theoretical works\cite{singh,dugaev} and a few reports 
on experiments,\cite{kobayashi,raffy,rubinstein,aprili,aliev,smorchkowa} 
which prove the existence and
importance of the quantum corrections related
to both weak localization and electron-electron interaction effects.
The theoretical description, however, was restricted 
to the effects of localization
on the spin-density fluctuations in the vicinity of the ferromagnetic 
transition\cite{singh} and to electron-electron interaction effects
in spin dependent quantum wells\cite{dugaev}.

It is well known that the quantum corrections due to weak localization
in nonmagnetic systems are suppressed by a sufficiently large
magnetic induction ${\bf B}$. One may then expect a similar suppression
of weak localization by an internal magnetic induction $B_{int}$
in ferromagnets.
But this point is still not definitely clear from the experimental point of view, 
at least for some kinds of
ferromagnetic materials. It is then reasonable to assume
that the internal magnetic induction existing inside the ferromagnets
may reduce the localization corrections instead of
destroying them totally. Very likely, one can expect only
a slight effect of $B_{int}$ in the case
of newly developed magnetic semiconductors like GaMnAs
alloys.\cite{ohno}

Spin-orbit (SO) scattering from paramagnetic impurities in nonmagnetic
metals is known to have a significant influence on the quantum corrections.
It can reverse the sign of the localization correction 
(so-called weak antilocalization effect), which results in a positive
magnetoresistance at weak
magnetic fields.\cite{anderson,hikami,maekawa}
However, SO interaction may also result from 
other sources, like for example the 
Dresselhaus\cite{dressel} or Bychkov-Rashba\cite{rashba} terms in the 
relevant Hamiltonian. These terms are related to the lack of inversion
symmetry in certain crystals or to reduced dimensionality
in quantum-confined structures,
respectively. In the context of weak localization theory, this type of 
spin-orbit interaction has been studied by A.~G.~Aronov and 
Yu.~B.~Lyanda-Geller.\cite{aronov,knap,lyanda98}

Recently, there is a large interest in SO interaction 
in magnetically ordered materials. First, the SO scattering is 
believed to be responsible for the anomalous Hall effect in 
ferromagnets.\cite{hirsch,bulgakov,zhang} Second, the SO interaction is
one of the main interactions which determine the spin diffusion length.
The latter quantity plays a crucial role in the dependence of the
giant magnetoresistance effect on the sublayer thicknesses, when the
current flows perpendicularly to the films.\cite{fert,bas}

In this paper we study the localization corrections
to conductivity of ferromagnetic systems in
the presence of SO scattering from defects and also in the
presence of the Bychkov-Rashba term. It is well known 
that in nonmagnetic materials the spin-orbit scattering is crucial
for the localization correction. As it was already mentioned above,
the SO scattering leads to weak antilocalization, i.e.,
to positive magnetoresistance at small magnetic
fields.\cite{anderson,hikami,maekawa} The situation in ferromagnetic
metals, however, is significantly different.
We show that the processes, leading to weak antilocalization in nonmagnetic 
systems, are totally suppressed in ferromagnets, so in the presence of SO 
interaction we have only a negative magnetoresistance.

In Sections 2 and 3 we derive the formula for the Cooperon
and quantum correction to conductivity in three-dimensional (3D)
ferromagnets. Two dimensional (2D) ferromagnets are considered in Section 4,
while quantum wells and possible crossover from 2D to 3D case are
discussed in Section 5. The influence of SO interaction in the form of
Bychkov-Rashba term is studied in Section 6. In Section 7 we discuss the 
effect of internal magnetization on the localization corrections. 
Finally, conclusions and summary are in Section 8.

\section{Cooperon in a 3D ferromagnet}

We consider the following model Hamiltonian of a ferromagnet
with SO scattering: 
$$ 
H=\int d^3{\bf r}\,  \psi ^{\dag }\left ({\bf r} \right) \left[
-\frac{\nabla^{2}}{2m}-M\sigma_{z}
+V\left( {\bf r}\right) \right] 
\psi \left( {\bf r} \right) \; ,
\eqno (1) 
$$ 
where the axis $z$ is assumed to be along the
magnetization ${\bf M}$, 
$\psi$ is a spinor field with the components $\psi_\uparrow$
and $\psi_\downarrow$, and we put $\hbar =1$. 
In the presence of a magnetic induction ${\bf B}={\rm rot}\, {\bf A}$, the
gradient operator $\nabla $ is replaced by $\nabla -i\, e{\bf A}/c$. Note that
the exchange term $-M\sigma _z$ acts only on the spin and has therefore 
no direct effect on the orbital motion of the electrons.

The random potential $V({\bf r})$ of impurities consists of two statistically
independent components -- the component independent of the electron spin,
described by the random potential
$V_0({\bf r})$, and the 
spin-orbit component $V_{so}({\bf r})$.
Matrix elements of the latter component have the form
$$
(V_{so})_{{\bf k}\alpha,{\bf k}'\beta}=i\, V_1\left( {\bf k}\times
{\bf k}'\right) \cdot 
{\bf \sigma}_{\alpha \beta } \; 
\eqno (2)
$$
for the transitions
$\left( {\bf k},\alpha \right) \rightarrow 
\left( {\bf k}',\beta\right) $,  
where $V_1$ is a constant, 
${\bf k}$ and ${\bf k}'$ are the initial and final electron wavevectors,
$\alpha$ and $\beta$ describe the corresponding spin states,
and ${\bf \sigma}=(\sigma_x, \sigma_y,\sigma_z)$ are the Pauli matrices.

The key component of the weak localization theory is the 
Cooperon,\cite{alt82,lee85,alt85,bergmann}
which can be presented by a ladder in the particle-particle channel with 
two propagators describing electrons with almost vanishing total momentum 
and with very close energy parameters (so-called Cooper channel). 
In the case of ferromagnets, and as long as
$M\gg \tau _\uparrow ^{-1},\tau _\downarrow ^{-1}$, where 
$\tau _\uparrow $ and $\tau _\downarrow  $ are the momentum 
relaxation times of the spin up and down electrons at the Fermi surface, 
this channel does not include ladder elements with Green functions 
corresponding to the opposite spin orientations. Indeed, using the standard
method of calculation of the Cooperon,\cite{alt82,alt85} one should evaluate
the integral
$$
\Pi _{\sigma \sigma ^\prime }=
\int \frac{d^3{\bf k}}{(2\pi )^3}\,  
G_\sigma ^R(\varepsilon ,{\bf k})\, G_{\sigma ^\prime }^A(\varepsilon, -{\bf k}), 
$$
where $G_\sigma ^{R,A}(\varepsilon ,{\bf k})$ are the retarded and advanced
Green functions of an electron 
with spin $\sigma =\uparrow ,\downarrow $. This gives 
$\Pi _{\uparrow \downarrow }/\Pi _{\uparrow \uparrow }\simeq 1/M\tau _\uparrow \ll 1 $,
and $\Pi _{\uparrow \downarrow }/\Pi _{\downarrow \downarrow }\simeq 1/M\tau _\downarrow \ll 1 $,
which corresponds to vanishingly small contribution of the singlet channel.
This result can also be understood as a suppression of singlet pairs by
the exchange field.

Validity of this approach is  confirmed by the following estimations. If
we assume parameters typical for pure Fe,\cite{hood} i.e.,  $M=2.5$~eV and 
$\tau _\uparrow \simeq \tau _\downarrow \simeq 5\times 10^{-13}$~s,
we obtain $(M\tau _{\uparrow ,\downarrow })^{-1}\simeq 5\times 10^{-4}$.  
In dirty Fe, this value would be increased by one or two orders of magnitude, 
but will still remain small as compare to unity.
The exclusion of the Cooperon in the singlet channel
is the crucial point of our description,
which leads to the absence of weak antilocalization in ferromagnets. 
In the following, we will omit the spin index in the
inequalities
$M\gg \tau _\uparrow ^{-1},\tau _\downarrow ^{-1}$, and will write simply
$M\gg \tau ^{-1}$. 

\begin{figure} 
\hskip2cm  
\epsfig{file=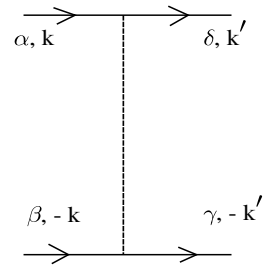,height=4cm,width=4cm} 
\caption{The bare scattering amplitude. }
\end{figure}

In the case of weak scattering potential and upon averaging over
the random field $V({\bf r})$, one finds the following bare scattering
amplitude (see Fig.~1): 
$$
\Gamma^{0}_{\alpha\beta\gamma\delta}
=a\delta_{\alpha\delta}\delta_{\beta\gamma}
-b\left( \sigma_{\alpha\delta}^{x}\sigma_{\beta\gamma}^{x}
+\sigma_{\alpha\delta}^{y}\sigma_{\beta\gamma}^{y}\right) 
-t_{\alpha \beta \gamma \delta  } \; ,
\eqno (3)
$$ 
where $t_{\alpha\beta \gamma\delta}$ is a diagonal matrix in the 
space of  $(\uparrow \uparrow ),(\uparrow \downarrow ),
(\downarrow \uparrow ),(\downarrow \downarrow )$ states,
$$
{\bf t}={\rm diag}\left( d_1,\, d_3,\, d_3,\, d_2\right) \; ,
\eqno (4)
$$
and we introduced the following definitions:
$$ 
a=N_i\, V_0^2
=\frac{1}{2\pi \nu _\uparrow \tau _{0\uparrow }}
=\frac{1}{2\pi \nu _\downarrow \tau _{0\downarrow }}\; , 
\eqno (5)
$$
$$
b=N_i\, V_1^2\, \lambda _0\, 
k_{F\uparrow }^2\, k_{F\downarrow }^2\; ,
\eqno (6)
$$
$$
d_1=N_i\, V_1^2\, \lambda _0\, k_{F\uparrow }^4 
=\frac{1}{2\pi \nu _\uparrow \tau _{so\uparrow }^z}\; , 
\eqno (7)
$$
$$
d_2=N_i\, V_1^2\, \lambda _0\, k_{F\downarrow }^4 
=\frac{1}{2\pi \nu _\downarrow \tau _{so\downarrow }^z}\; , 
\eqno (8)
$$
$$
d_3=-N_i\, V_1^2\, \lambda _0\, 
k_{F\uparrow }^2\, k_{F\downarrow }^2\; ,
\eqno (9)
$$
with $\lambda_0$ defined as 
$$
\lambda _0=\overline{\left( {\bf n}_{\bf k} 
\times {\bf n}_{\bf k'}\right) _x^2}=\frac29 \; .
\eqno (10)
$$
In the above equations $N_i$ is the concentration of scattering centers,
$\nu _{\uparrow }$ and $\nu _{\downarrow }$ 
are the densities of states at the Fermi level for spin-up (majority) and
spin-down (minority) electrons, while $k_{F\uparrow}$
and $k_{F\downarrow}$ are the Fermi wavevectors for
spin-up and spin-down electrons.
The relaxation times $\tau _{0\uparrow}$
and  $\tau _{0\downarrow }$,  
defined by Eq.~(5), are the momentum relaxation 
times in the absence of SO scattering, whereas the relaxation times  
$\tau _{so\uparrow }^z$ and $\tau _{so\downarrow }^z$,  
defined by Eqs.~(7) and (8), are due to the SO scattering.
The averaging in Eq.~(10) is over
all orientations of the unit vectors ${\bf n}_{\bf k}$ and 
${\bf n}_{\bf k'}$, where ${\bf n}_{\bf k}={\bf k}/k$
and ${\bf n}_{\bf k'}={\bf k}'/k'$. 

Using Eq.~(3), we find the following
bare scattering amplitudes for the spin-up and spin-down electrons:
$$
\Gamma _\uparrow ^0
\equiv \Gamma _{\uparrow\uparrow\uparrow\uparrow}^0=a-d_1
=\frac{1}{2\pi \nu _\uparrow \tau _{0\uparrow }}
-\frac{1}{2\pi \nu _\uparrow \tau _{so\uparrow }^z}\; ,  
\eqno (11)
$$
$$
\Gamma _\downarrow ^0
\equiv \Gamma _{\downarrow\downarrow\downarrow\downarrow}^0=a-d_2
=\frac{1}{2\pi \nu _\downarrow \tau _{0\downarrow }}
-\frac{1}{2\pi \nu _\downarrow \tau _{so\downarrow }^z}\; .  
\eqno (12)
$$
One should note at this point that the bare elements
$\Gamma _{\uparrow\downarrow\uparrow\downarrow}^0$ and 
$\Gamma _{\downarrow\uparrow\downarrow\uparrow}^0$
do not contribute to the Cooper-channel diagrams
in the case of ferromagnetic systems as long as
$M>>1/\tau$, as we already stated before.  Apart from this, 
in a 3D case $\Gamma ^0$ does not contain the components 
$\Gamma _{\uparrow\uparrow\downarrow\downarrow}^0$ and 
$\Gamma _{\downarrow\downarrow\uparrow\uparrow}^0$, which 
vanish due to the rotational symmetry in the $x$-$y$ plane, 
as can also be concluded from Eq.~(3). 

\begin{figure}  
\epsfig{file=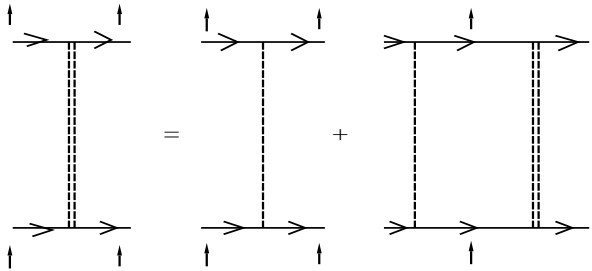,height=5cm,width=9cm} 
\caption{Ladder diagram for the Cooperon in 3D case. }
\end{figure}

Summing up the ladder diagrams
contributing to the renormalized scattering
amplitude $\Gamma _\uparrow (\omega ,q)\equiv
\Gamma _{\uparrow\uparrow\uparrow\uparrow}(\omega ,q)$  
in the Cooper channel
with small transferred energy $\omega $ and momentum
$q\; $\cite{alt82,lee85} (see Fig.~2), one
obtains the equation
$$
\Gamma _\uparrow (\omega ,q) =\Gamma _\uparrow ^0
+\Gamma _\uparrow ^0\, \Pi _\uparrow (\omega ,q)\, 
\Gamma _\uparrow (\omega ,q)\; ,
\eqno (13)
$$
where
$$
\Pi _\uparrow (\omega ,q)
=\int \frac{d^3{\bf k}}{(2\pi )^3}
G_\uparrow ^R\left( \omega ,{\bf k}+{\bf q} \right) \,   
G_\uparrow ^A\left(0,\, -{\bf k}\right)
$$
$$
\simeq 2\pi \nu _\uparrow \tau _\uparrow \,
\left( 1-D_\uparrow q^2\tau _\uparrow 
+i\omega \tau _\uparrow \right) \; .
\eqno (14)
$$
Here, $D_\uparrow =\frac13 \, v_{F\uparrow }^2\tau _\uparrow $ is the
diffusion coefficient, $v_{F\uparrow }$ is 
the Fermi velocity, and $\tau _\uparrow $ is the momentum 
relaxation time of spin up electrons. 
Equation (14) was derived in the diffusion limit, i.e., when 
$\omega \ll 1/\tau _\uparrow $ and 
$q\ll (D_\uparrow \tau _\uparrow )^{-1/2}$.  

\begin{figure} 
\epsfig{file=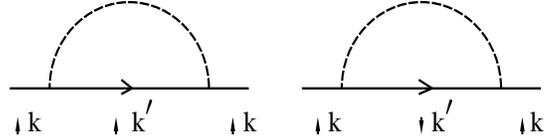,height=5cm,width=8.5cm} 
\caption{Self energy diagram. }
\end{figure}

The relaxation time $\tau _\uparrow $ can be found by calculating the
self energy, presented by the diagrams of Fig.~3. The self energy contains 
non-zero spin-flip vertices of the singlet type, as shown in Fig.~4.
After calculating the imaginary part of the self energy, we find
$$
\frac1{\tau _\uparrow }=2\pi \nu _\uparrow \left( 
a+d_1+\frac{2\nu _\downarrow }{\nu _\uparrow }\, b\right) \; .
\eqno (15)
$$  

\begin{figure} 
\hskip1cm
\epsfig{file=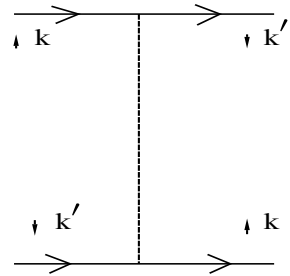,height=5cm,width=5cm} 
\caption{Spin-flip vertex. }
\end{figure}

Using Eqs.~(5) to (14), we finally obtain the following formula for
the renormalized scattering amplitude
in the Cooper channel (Cooperon):
$$
\Gamma _\uparrow (\omega ,q) 
=\frac{1}{\tau _\uparrow }\;
\frac{a+d_1+\left( 2\nu _\downarrow /\nu _\uparrow \right) b}
{-i\omega +D_\uparrow q^2+1/\tilde{\tau }_{so\uparrow }}\; ,
\eqno (16)
$$
where the effective SO relaxation time of Cooperon,
$\tilde{\tau }_{so\uparrow }$, is introduced,
$$
\frac{1}{\tilde{\tau }_{so\uparrow }}
=\frac{2}{\tau _\uparrow }\;
\frac{d_1+\left( \nu _\downarrow /\nu _\uparrow \right) b}
{a-d_1}\; .
\eqno (17)
$$
Let us now  define
$\tau _{so\uparrow }^x$ and $\tau _{so\downarrow }^x$ as
$1/\tau _{so\uparrow }^x=2\pi \nu _\uparrow b$
and $1/\tau _{so\downarrow }^x=2\pi \nu _\downarrow b$.
In the limit of weak SO scattering, 
$\tau _{so\uparrow }^z\gg \tau _\uparrow $, one may then write
$$
\Gamma _\uparrow (\omega ,q)
=\frac{1}{2\pi \nu _\uparrow \tau _\uparrow ^2}\;
\frac{1}{-i\omega +D_\uparrow q^2+1/\tilde{\tau }_{so\uparrow }
+1/\tau _{\varphi \uparrow }}\; ,
\eqno (18)
$$
where
$$
\frac1{\tilde{\tau }_{so\uparrow }}
=2\left( \frac1{\tau _{so\uparrow }^z}
+\frac{\nu _\downarrow }{\nu _\uparrow }
\frac1{\tau _{so\uparrow }^x} \right)\; , 
\eqno (19)
$$
and a phase relaxation time $\tau _{\varphi \uparrow }$, related 
to some inelastic scattering processes of
electrons,\cite{alt82,lee85,alt85}
is added into Eq.~(18).  

The analogous formulae can also be derived for the spin-down Cooperon
$\Gamma _\downarrow (\omega ,q)\equiv
\Gamma_{\downarrow\downarrow\downarrow\downarrow} (\omega ,q)$.
This formula can be obtained from Eqs. (18) and (19) by inverting 
direction of the arrows.

\section{Conductivity and Magnetoconductivity in 3D case}

The localization correction $\Delta\sigma$ to the static
conductivity is determined by the
diagrams\cite{alt82,lee85} shown in Fig.~5.

\begin{figure} 
\epsfig{file=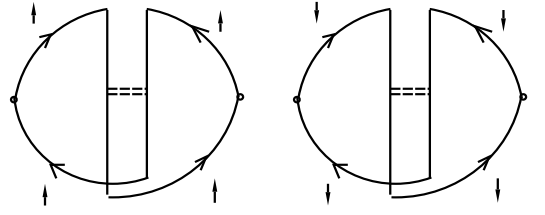,height=5cm,width=8.5cm} 
\caption{Localization corrections to conductivity. }
\end{figure}

Upon calculating the corresponding
contributions, one finds the following formula for $\Delta \sigma$:
\wide{m}{
$$
\Delta \sigma 
=-\frac{e^2}{\pi }
\left( 2\pi \nu _\uparrow \tau _\uparrow ^2 D_\uparrow 
\int \frac{d^3{\bf q}}{(2\pi)^3} \Gamma _\uparrow (0,q)
+2\pi \nu _\downarrow \tau _\downarrow ^2 D_\downarrow 
\int \frac{d^3{\bf q}}{(2\pi)^3}2\pi \nu _\downarrow \Gamma _\downarrow (0,q) 
\right) \; ,
\eqno (20)
$$
which is a straightforward generalization of the localization correction
in a nonmagnetic case.
Using Eqs. (18) and (20), we obtain
$$
\Delta \sigma ={\rm const}
+\frac{e^2}{4\pi ^2}\left[
\frac{1}{D_\uparrow ^{1/2}}
\left( \frac{1}{\tilde{\tau }_{so\uparrow }}
+\frac{1}{\tau _{\varphi \uparrow }}\right) ^{1/2}
+\frac{1}{D_\downarrow ^{1/2}}
\left( \frac{1}{\tilde{\tau }_{so\downarrow}}
+\frac{1}{\tau _{\varphi \downarrow }}\right) ^{1/2}\right] \; .
\eqno (21)
$$
where the constant part is related to the contribution from the largest
momenta of Cooperon, $q\sim (D\tau )^{-1/2}$, and can not be calculated
exactly within the diffusion approximation.\cite{alt82,alt85} 
It can be estimated as 
$$
{\rm const}\simeq-\frac{e^2}{4\pi ^2}\left[ (D_\uparrow \tau _\uparrow )^{-1/2}
+(D_\downarrow \tau _\downarrow )^{-1/2}\right] .
$$
Thus, the constant term is negative and, since
$\tau _{\uparrow ,\downarrow }<\tilde{\tau }_{so},\tau _\varphi $,
it is larger in magnitude than the second term of Eq.~(21). 
Therefore, the total correction (21) is negative.
By decreasing $\tilde{\tau} _{so}$ and/or $\tau _\varphi $, we suppress the 
localization correction to conductivity.

The magnetic induction (both external and internal) supresses the localization
corrections. If the magnitude of the total magnetic induction is $B$, then,
following the method developed by Kawabata,\cite{kawabata} we find    
$$
\Delta \sigma (B)
=-\frac{e^2}{\pi }
\left( D_\uparrow \frac{eB}{2\pi c} \sum _{n=0}^{n_{0\uparrow }} 
\int _{-\infty }^{\infty}\frac{dq}{2\pi}\left[
D_\uparrow q^2 +\frac{4eBD_\uparrow }{c}\left( n+\frac12 \right)
+\frac1{\tilde{\tau }_{so\uparrow }}
+\frac1{\tau _{\varphi \uparrow }}\right]^{-1}\right.
$$
$$
\left.
+D_\downarrow \frac{eB}{2\pi c} \sum _{n=0}^{n_{0\downarrow }} 
\int _{-\infty }^{\infty}\frac{dq}{2\pi}\left[
D_\downarrow q^2 +\frac{4eBD_\downarrow }{c}\left( n+\frac12\right) 
+\frac1{\tilde{\tau }_{so\downarrow }}
+\frac1{\tau _{\varphi \downarrow }}\right]^{-1} \right) \; ,
\eqno (22)
$$
where $l_B=(c/eB)^{1/2}$ is the magnetic length, $c$ is the light velocity,
and the sums over the
Landau levels are cut off at 
$n_{0\uparrow (\downarrow )} 
\simeq l_B^2/\left( D_{\uparrow (\downarrow )}
\tau _{\uparrow (\downarrow )}\right) $. 

After eliminating the $\Delta\sigma (B=0)$ part,\cite{kawabata} we find a formula, 
which is a generalization of the  Kawabata's low-field magnetoresistance to the 
ferromagnetic case,
$$
\Delta \sigma (B)-\Delta \sigma (0)=
-\frac{e^2}{16\pi ^2l_B}\sum _{n=0}^\infty \left [
\frac1{(n+1/2+\delta _\uparrow )^{1/2}}
-2(n+1+\delta _\uparrow )^{1/2}+2(n+\delta _\uparrow )^{1/2}\right.
$$
$$
\left.  
+\frac1{(n+1/2+\delta _\downarrow )^{1/2}}
-2(n+1+\delta _\downarrow )^{1/2}
+2(n+\delta _\downarrow )^{1/2}\right] \; , 
\eqno (23)
$$
}
where
$$
\delta _{\uparrow (\downarrow )}
=\frac{l_B^2}{4D_{\uparrow (\downarrow )}}
\left( \frac1{\tilde{\tau }_{so\uparrow (\downarrow )}}
+\frac1{\tau _{\varphi \uparrow (\downarrow )}} \right)\; .
\eqno (24)
$$
In accordance with Eq.~(22), the magnetic induction supresses the negative
correction to the conductivity. Thus, the resulting sign of Eq.~(23) is positive,
and its magnitude, $\Delta \sigma (B)-\Delta \sigma (0)$, increases with increasing 
magnetic field.
This means that one finds a negative magnetoresistance, despite of the presence 
of  the spin-orbit interaction. The reason of this is the fact that we have
excluded the singlet Cooperon, which contributes to the localization
correction with the opposite sign, and usually gives rise to
a positive magnetoresistance (weak antilocalization) in weak magnetic fields in
nonmagnetic materials.\cite{alt82,lee85,bergmann} 

The obtained result is in agreement with the results on magnetoconductivity of
nonmagnetic metals in a magnetic field, if both Zeeman splitting and
spin-orbit scattering are taken into account.\cite{maekawa,alt85} 
Indeed, the exchange field of a ferromagnet enters the Hamiltonian, Eq.~(1),
like the Zeeman term. Thus, the strong exchange-field limit of a ferromagnet 
corresponds to the case of a large magnetic field in the Zeeman term. 
At these conditions, the effect of a magnetic field is associated with a negative 
magnetoresistance due to the suppression of the singlet Cooperon
by magnetic field through the Zeeman splitting.

\section{Two-dimensional ferromagnets}

In this section we consider a two-dimensional ferromagnet. In such a case  
there is no electron motion in the direction perpendicular
to the plane, and consequently the electron wavevectors are in
the plane of the ferromagnet.
We consider first the case of in-plane magnetization,
as shown schematically in Fig.~6.

\begin{figure} \hskip1cm
\epsfig{file=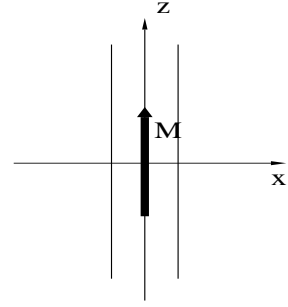,height=5cm,width=5cm} 
\caption{2D ferromagnet with in-plane magnetization. }
\end{figure}

The bare scattering amplitude has then the form
$$
\Gamma^{0}_{\alpha\beta\gamma\delta}
=a\, \delta_{\alpha\delta}\delta_{\beta\gamma}
-b\, \sigma_{\alpha\delta}^{x}\sigma_{\beta\gamma}^{x}\; ,
\eqno (25)
$$ 
where
$$
a=\frac1{2\pi \nu \tau _0} 
\eqno (26)
$$
and
$$
b=\frac1{2\pi \nu \tau _{so}^x}\; .
\eqno (27)
$$
In a strictly 2D case and when both spin sub-bands are populated with
electrons, the density of states is independent of the
spin orientation, $\nu_\uparrow =\nu_\downarrow\equiv\nu$.
This follows from the model assumed in Eq.~(2). 
In that case also
the relaxation times in the absence and/or presence of the SO scattering
are independent of the electron spin,  
$\tau_{0\uparrow}=\tau_{0\downarrow}\equiv\tau_0$
and $\tau_{so\uparrow}^x=\tau_{so\downarrow}^x\equiv\tau_{so}^x$.

Contrary to the 3D case considered above, we have now to include
the spin-flip processes in the Cooperon ladder. This is due to the 
fact that now we do not have the rotational symmetry
in the  $x$-$y$ plane. According to Eqs.~(25) to (27), the bare
amplitudes are now,
$$
\Gamma ^0_\uparrow =\Gamma ^0_\downarrow 
\equiv \Gamma ^0=\frac1{2\pi \nu \tau _0}\; ,
\eqno (28)
$$
$$
\Gamma ^0_{\uparrow\uparrow\downarrow\downarrow}
=\Gamma ^0_{\downarrow\downarrow\uparrow\uparrow} 
\equiv \Gamma _{sf}^0=-\frac1{2\pi \nu \tau _{so}^x}\; .
\eqno (29)
$$
The equations for the renormalized vertices can be written as two
coupled ladders for $\Gamma_{\uparrow\uparrow\uparrow\uparrow}(\omega ,q)=
\Gamma_{\downarrow\downarrow\downarrow\downarrow}(\omega ,q)\equiv
\Gamma(\omega ,q) $ and $\Gamma_{\uparrow\uparrow\downarrow\downarrow}(\omega ,q)=
\Gamma_{\downarrow\downarrow\uparrow\uparrow}(\omega ,q)\equiv
\Gamma _{sf}(\omega ,q)$, as shown in Fig.~7.

\begin{figure} 
\epsfig{file=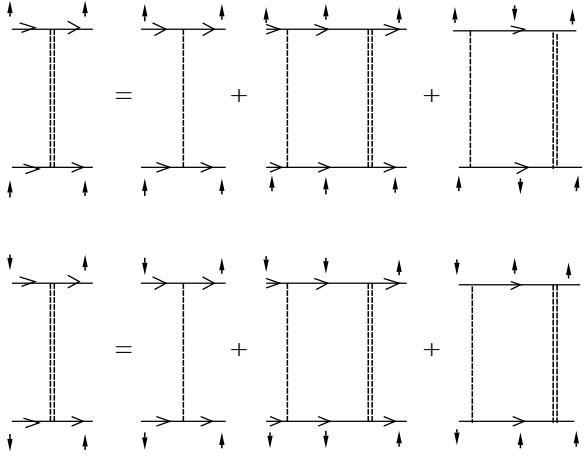,height=7cm,width=9cm}
\vskip0.5cm 
\caption{Equation for the Cooperon in 2D ferromagnet with in-plane
magnetization. }
\end{figure}

From these equations we find
$$
\Gamma (\omega ,q)=\frac1{4\pi \nu \tau ^2}\,
\frac1{-i\omega +\overline{D}q^2+1/\tau _\varphi }\; ,
\eqno (30)
$$
where $\overline{D}=\frac12\left( D_\uparrow +D_\downarrow \right)$,
and the diffusion constants $D_\uparrow$ and $D_\downarrow$ are 
defined as
$D_{\uparrow (\downarrow )}=\frac12\, v_{F\uparrow (\downarrow )}^2\tau $. 
The electron relaxation time $\tau$ in Eq.~(30) is independent
of the electron spin, $\tau_\uparrow =\tau_\downarrow \equiv \tau$, and can  
be calculated in the same way as in the 3D case, which gives
$$
\frac1{\tau_\uparrow}=\frac1{\tau_\downarrow}\equiv\frac1{\tau }=
\frac1{\tau _0}
+\frac1{\tau _{so}^x}\; .
\eqno (31)
$$
It is worth to note that the spin-orbit scattering enters the Cooperon only
through the one-particle relaxation time $\tau$, and has no influence on
the pole of the Cooperon.

\begin{figure} \hskip1cm
\epsfig{file=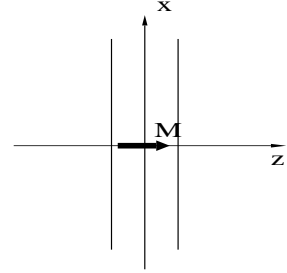,height=4.5cm,width=5cm} 
\caption{2D ferromagnet with magnetization perpendicular to its plane. }
\end{figure}

Consider now the case when the magnetization field ${\bf M}$ 
is perpendicular to the plane of the ferromagnet, as shown schematically
in Fig.~8.
The calculations are similar to those in the case of in-plane magnetization,
so we write down only the results.
The only bare scattering amplitudes are now
$\Gamma _{\uparrow}^0$ and $\Gamma _{\downarrow}^0$, which 
generally are different and have the form  
$$
\Gamma _{\uparrow (\downarrow)}^0=\frac1{2\pi \nu \tau _0}
-\frac1{2\pi \nu \tau _{so\uparrow (\downarrow) }^z}\; .
\eqno (32)
$$
The Cooperons  
$\Gamma _{\uparrow }(\omega ,q)$ and $\Gamma _{\downarrow}(\omega ,q)$
can be written as
$$
\Gamma _{\uparrow  (\downarrow)}(\omega ,q)
=\frac1{2\pi \nu \tau _{\uparrow  (\downarrow)}^2}\,
\frac1{-i\omega +D_{\uparrow (\downarrow)} q^2
+1/\tilde{\tau }_{so\uparrow  (\downarrow)}
+1/\tau _{\varphi\uparrow (\downarrow )}}\; ,
\eqno (33)
$$
where the relaxation times  $\tau_\uparrow$ and $\tau_\downarrow$
are given by 
$$
\frac1{\tau _{\uparrow (\downarrow)}}=\frac1{\tau _0}
+\frac1{\tau _{so\uparrow (\downarrow) }^z}\; ,
\eqno (34)
$$
and 
$$
\frac1{\tilde{\tau }_{so{\uparrow  (\downarrow)}}}
=\frac2{\tau _{so\uparrow  (\downarrow)}^z-2\tau _{\uparrow (\downarrow)} }
\simeq \frac2{\tau _{so\uparrow (\downarrow) }^z}\; ,
\eqno (35)
$$
provided that $\tau _{so\uparrow (\downarrow )}^z\gg
\tau _{\uparrow (\downarrow )}$.
By comparing the results (30) and (33), both obtained for  a 
2D case but for different magnetic configurations,
we see that the effect of SO interaction 
significantly depends on the orientation of the magnetization field 
${\bf M}$ with respect to the plane of the ferromagnet.

The quantum correction to the conductivity of a 2D ferromagnet takes
the form
\wide{m}{
$$
\Delta \sigma =\frac{e^2}{4\pi ^2}\left\{
\ln \left[ \tau _\uparrow \left( 
\frac1{\tau _{\varphi \uparrow }}
+\frac1{\tilde{\tau }_{so\uparrow }}\right) \right]
+\ln \left[ \tau _\downarrow \left( 
\frac1{\tau _{\varphi \downarrow }}
+\frac1{\tilde{\tau }_{so\downarrow }}\right) \right] \right\}\; ,
\eqno (36a)
$$
}
which is a direct generalization of the corresponding
formula in the 2D nonmagnetic case.\cite{alt82,lee85,bergmann}
Here,  $\tau_{\uparrow(\downarrow)} $ and
$\tilde{\tau }_{so\uparrow (\downarrow)}$
are defined, respectively, by (31)
and $\tilde{\tau }_{so\uparrow (\downarrow)}=0$ for the in-plane
magnetization, and by Eqs.~(34) and (35) for the
case of perpendicular magnetization.

The 2D localization correction, described  by Eq.~(36a), is negative 
since $\tau <\tilde{\tau }_{so},\tau _\varphi $, and, in addition, 
we take here $\tau \ll \tilde{\tau} _{so},\tau _\varphi $. The latter inequality means
that the momentum relaxation time of electrons,  $\tau $, is mainly due to the potential
scattering. 

We can also present an expression for the conductivity in the case of nonzero
magnetic induction ${\bf B}$, perpendicular to the plane, by generalizing the 
result for a nonmagnetic two-dimensional system\cite{hikami}
$$
\Delta \sigma (B)=-\frac{e^2}{4\pi ^2}\left[
\psi \left( \frac12+\frac1{\tau _\uparrow a_\uparrow }\right)
-\psi \left( \frac12+\frac1{\tilde{\tau }_{so \uparrow }a_\uparrow }
+\frac1{\tau _{\varphi \uparrow }a_\uparrow }\right)
\right.
$$
$$
\left.
+\psi \left( \frac12+\frac1{\tau _\downarrow a_\downarrow }\right)
-\psi \left( \frac12+\frac1{\tilde{\tau }_{so \downarrow }a_\downarrow }
+\frac1{\tau _{\varphi \downarrow }a_\downarrow }\right)
\right] ,
\eqno (36b)
$$   
where $a_{\uparrow ,\downarrow }=4eBD_{\uparrow, \downarrow }/c$,
and $\psi (x)$ is the Digamma function,\cite{abramowitz} which has the property
$\psi (x)\simeq \ln (x) $ for $x\gg 1$. 

The magnetic induction suppresses the negative correction to the conductivity, 
which leads to the negative magnetoresistance.
It should be noted that in strongly 2D case, the in-plane magnetic induction does 
not affect the localization correction to conductivity. The reason is that in
the two-dimensional case, the flux of magnetic induction does not penetrate 
through any closed electron paths. Correspondingly, the in-plane induction
does not break the interference of closed trajectories of electrons
moving in opposite directions, responsible for the weak localization 
effect.\cite{alt82,lee85,alt85,bergmann}

\section{Quantum wells}

In a quasi 2D case the electrons are confined within a quantum
well and the number $N_F$ of 2D subbands populated with electrons
is larger than one, $N_F>1$.
The situation with a large number of occupied subbands 
is typical for metals, whereas the situation with only a few
populated subbands is characteristic of semiconductor quantum wells.
The effect of SO scattering
on the magnetoresistance of nonmagnetic materials at such conditions
has been studied by Hikami et al.\cite{hikami}

We consider here a ferromagnetic quantum well in the geometry
shown in Fig.~6, i.e., when  
the magnetization field ${\bf M}$ is in the film plane.
The scattering amplitude has then the following form
$$
\Gamma^{0}_{\alpha\beta\gamma\delta}
=a\, \delta_{\alpha\delta}\delta_{\beta\gamma}
-b\, \sigma_{\alpha\delta}^{x}\, \sigma_{\beta\gamma}^{x}
-c\, \sigma_{\alpha\delta}^{y}\, \sigma_{\beta\gamma}^{y} 
-t_{\alpha \beta \gamma \delta  } \; ,
\eqno (37)
$$ 
with $a$, $b$ and  $t_{\alpha\beta \gamma\delta }$ defined by
Eqs.~(4) to (6), and  
$$
c=N_i\, V_1^2\,
\overline{\left( {\bf k}
\times {\bf k}'\right) _y^2}^{(\uparrow ,\downarrow )}
=\frac{1}{2\pi \nu _\uparrow \tau _{so\uparrow }^y} 
=\frac{1}{2\pi \nu _\downarrow \tau _{so\downarrow }^y}\; , 
\eqno (38)
$$
$$
d_1=N_i\, V_1^2\, 
\overline{\left( {\bf k}
\times {\bf k}'\right) _z^2}^{(\uparrow )}
=\frac{1}{2\pi \nu _\uparrow \tau _{so\uparrow }^z}\; , 
\eqno (39)
$$
$$
d_2=N_i\, V_1^2\,  
\overline{\left( {\bf k}
\times {\bf k}'\right)_z^2}^{(\downarrow )}
=\frac{1}{2\pi \nu _\downarrow \tau _{so\downarrow }^z}\; , 
\eqno (40)
$$
$$
d_3=-N_i\, V_1^2\,  
\overline{\left( {\bf k}
\times {\bf k}'\right)_z^2}^{(\uparrow ,\downarrow )}\; .
\eqno (41)
$$
The averages in Eqs.~(38) to (41) include averaging over discrete 
subbands due to quantization of the electron motion
along the axis $x$ (normal to the film plane).
In this notation, the expressions for bare vertices
$\Gamma _\uparrow ^0$ and $\Gamma _\downarrow ^0$ coincide with 
Eqs. (11) and (12), whereas
for the spin-flip vertex $\Gamma _{sf}^0$ we have
$$
\Gamma _{sf}^0=-\frac1{2\pi \nu _\uparrow \tau _{so\uparrow }^x}
+\frac1{2\pi \nu _\uparrow \tau _{so\uparrow }^y} \; . 
\eqno (42)
$$

The ladder equations for
$\Gamma _\uparrow (\omega ,q)$, $\Gamma _\downarrow (\omega ,q)$
and $\Gamma _{sf}(\omega ,q)$
have the form shown in Fig.~7. From these equations one finds the
following solution for $\Gamma _{\uparrow }(\omega ,q)$:
\wide{m}{
$$
\Gamma _\uparrow (\omega ,q)
=\frac{(\Gamma _{sf}^0)^2\, \Pi _{\downarrow }(\omega ,q) 
+\Gamma _{\uparrow }^0 
\left[ 1-\Gamma _{\downarrow }^0\, \Pi _{\downarrow }(\omega ,q)\right] }
{\left[ 1-\Gamma _{\uparrow }^0\, \Pi _{\uparrow }(\omega ,q)\right]  
\left[ 1-\Gamma _{\downarrow }^0\, \Pi _{\downarrow }(\omega ,q)\right]
-(\Gamma _{sf}^0)^2\, \Pi _{\uparrow }(\omega ,q)\, \Pi _{\downarrow }(\omega ,q)}\; .
\eqno (43)
$$
In the case of weak SO interaction, 
the final expression for $\Gamma _{\uparrow }(\omega ,q)$ takes the 
form
$$
\Gamma _{\uparrow }(\omega ,q)
=\frac1{2\pi \nu _{\uparrow }\tau _\uparrow ^2}\,
\frac{A_\uparrow }{-i\omega +\widetilde{D}_\uparrow q^2
+1/\tilde{\tau }_{so\uparrow }+1/\tau _{\varphi \uparrow}}\; ,
\eqno (44)
$$
where
$$
A_\uparrow =\frac{2/\tau _{so\downarrow }^z
+1/\tau _{so\uparrow }^x+1/\tau _{so\uparrow }^y}
{2/\tau _{so\uparrow }^z+2/\tau _{so\downarrow }^z
+\left( 1/\tau _{so\uparrow }^x+1/\tau _{so\uparrow }^y\right)
\left( 1+\nu _{\downarrow }/\nu _{\uparrow }\right) }\; ,
\eqno (45)
$$ 
$$
\widetilde{D}_{\uparrow }
=\frac{D_\uparrow \left( 2/\tau _{so\downarrow }^z
+1/\tau _{so\uparrow }^x+1/\tau _{so\uparrow }^y\right) 
+D_\downarrow \left( 2/\tau _{so\uparrow }^z
+1/\tau _{so\downarrow }^x+1/\tau _{so\downarrow }^y\right) }
{2/\tau _{so\uparrow }^z+2/\tau _{so\downarrow }^z
+\left( 1/\tau _{so\uparrow }^x+1/\tau _{so\uparrow }^y\right)
\left( 1+\nu _{\downarrow }/\nu _{\uparrow }\right) }\; ,
\eqno (46)
$$ 
$$
\frac1{\tilde{\tau }_{so\uparrow }}
=2\, \frac{2/\left( \tau _{so\uparrow }^z\, \tau _{so\downarrow }^z\right) 
+\left( 1/\tau _{so\uparrow }^x+1/\tau _{so\uparrow }^y\right)
\left( 1/\tau _{so\uparrow }^z
+\nu _{\downarrow }/
\left( \nu _{\uparrow }\, \tau _{so\downarrow }^y\right) \right)
+2/\left( \tau _{so\uparrow }^x\, \tau _{so\downarrow }^y\right) }
{2/\tau _{so\uparrow }^z+2/\tau _{so\downarrow }^z
+\left( 1/\tau _{so\uparrow }^x+1/\tau _{so\uparrow }^y\right)
\left( 1+\nu _{\downarrow }/\nu _{\uparrow }\right) }\; ,
\eqno (47)
$$ 
}
and the spin-up relaxation time $\tau _{\uparrow }$ is
$$
\frac1{\tau _{\uparrow }}
=\frac1{\tau _{0\uparrow }}+\frac1{\tau _{so\uparrow }^z}
+\frac1{\tau _{so\downarrow }^x}+\frac1{\tau _{so\downarrow }^y}\; .
\eqno (48)
$$
The corresponding expression for $\Gamma _{\downarrow }(\omega ,q)$
can be obtained from Eqs.~(44) to (48) by changing the arrow direction
in these formulas. It is worth to note, that 
according to Eq.~(46), the spin-orbit interaction renormalizes the 
diffusion coefficients.

The correction to conductivity is described either by Eq.~(21) 
or by Eq.~(36), in dependence on the ratio of 
$L_{eff}=\left( \widetilde{D}/(\tilde{\tau }_{so}^{-1}
+\tau _{\varphi }^{-1})\right) ^{1/2}$ 
to the width of the
quantum well $L$. For $L_{eff}<L$ we have effectively a 3D case
described by Eq.~(21), whereas for $L_{eff}\gg L$ one finds effectively
2D behavior, Eq.~(36a).
It is worth to note that due to a strong asymmetry between spin-up and spin-down 
states, a "mixed" case is possible, with a 3D correction for one spin orientation 
and  a 2D correction for the opposite spin orientation.

\section{Spin-orbit interaction of the Bychkov-Rashba type}

The case when the spin-orbit coupling enters the Hamiltonian also in
the absence of scattering defects needs a special treatment.
In the following we will consider a system described by the Hamiltonian including
the Bychkov-Rashba
interaction term\cite{rashba}
$$
H=\int \frac{d^3{\bf k}}{(2\pi )^3}\; \psi_
{\bf k}^\dag \left[
\frac{k^2}{2m}-M\sigma _z
+g\left( k_y\sigma _z-k_z\sigma _y\right) \right]
\psi_{\bf k}\; ,
\eqno (49)
$$
where the magnetization $\bf M$ is assumed to be in 
the plane of the system (the geometry of Fig.~6) and $g\equiv g(k)$ 
is the spin-orbit coupling parameter.
This Hamiltonian can describe electrons
in low-dimensional structures close to interfaces,
or electrons in a quantum
well with variable doping. 

The energy spectrum of the  Hamiltonian (49) has two branches,
$$
\varepsilon _{1(2)}\left( {\bf k}\right)
=\frac{k^2}{2m}
\pm \left[ \left( M-gk_y\right) ^2+g^2k_z^2\right]^{1/2}\; ,
\eqno (50)
$$
which are no longer pure spin-up and spin-down states,
but correspond to spin-mixed states. 
The eigenfunctions corresponding to the eigenvalues (50)
can be written as
$$
|1{\bf k}\rangle =\left( 4M^2+g^2k_z^2\right) ^{-1/2}
\left( -igk_z|{\bf k}\uparrow \rangle
+2M|{\bf k}\downarrow \rangle\right)\; ,
\eqno (51)
$$
$$
|2{\bf k}\rangle =\left( 4M^2+g^2k_z^2\right) ^{-1/2}
\left( 2M|{\bf k}\uparrow \rangle
+igk_z|{\bf k}\downarrow \rangle \right)\; .
\eqno (52)
$$
Due to the terms linear in $k_y$ and for $M\ne 0$,
the energy spectrum
(50) is not symmetrical with respect to the 
${\bf k}\rightarrow -{\bf k}$ transformation,
as can also be seen in Fig.~9, (a) and (b).

\begin{figure} 
\epsfig{file=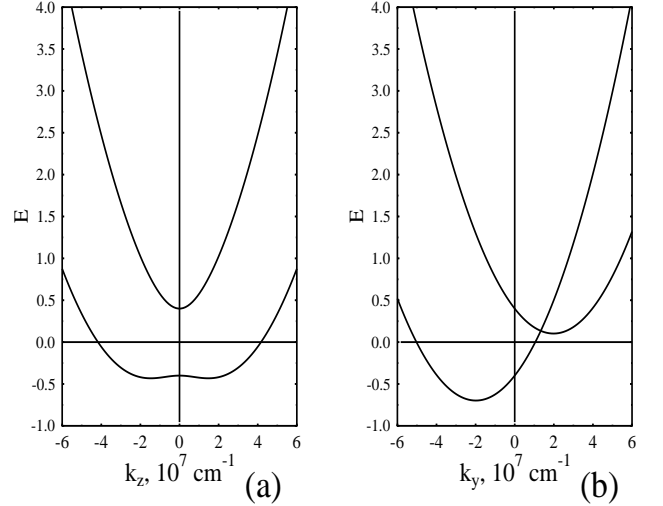,height=7cm,width=8.5cm} 
\caption{The energy spectrum E({\bf k}) of a ferromagnet with the
Bychkov-Rashba SO coupling, shown as a function of $k_z$ at 
$k_x=k_y=0$ (a) and as a function of $k_y$ at $k_x=k_z=0$ (b). }
\end{figure}

In the absence of external electric field $E$, there are nonvanishing
spin currents, associated with each branch of the spectrum
$\varepsilon _{1,2}({\bf k})$. However, it can be verified that the 
total charge current is zero if $E=0$. 

In the basis of the spin-up and spin-down states,
the electron Green function
has the matrix form,
$$
G_0\left( \varepsilon ,{\bf k}\right)
=\frac{\varepsilon -k^2/2m+\mu -M\sigma _z+gk_y\sigma _z-gk_z\sigma _y}
{\left( \varepsilon -\varepsilon _{1{\bf k}}
+\mu +i\delta \, {\rm sign}\, \varepsilon \right)
\left( \varepsilon -\varepsilon _{2{\bf k}}
+\mu +i\delta \, {\rm sign}\, \varepsilon \right) }\; ,
\eqno (53)
$$
where $\mu $ is the chemical potential. Accordingly, the self energy is 
also a matrix in this basis.
It is, however, more convenient to consider the electron self energy 
$\Sigma (\varepsilon ,{\bf k})$ in the basis 
of the eigenfunctions (51) and (52).
The self energy  is then diagonal and the imaginary
parts of $\Sigma _1(0, k_{F1})$ and $\Sigma _2(0, k_{F2})$
give the relaxation times $\tau _1$ and $\tau _2$
in the presence of defects.

Using Eqs.~(51) and (52), one can calculate the matrix elements of 
the impurity potential $V_0({\bf r})$,
$$
V_{1{\bf k},1{\bf k}'}=V_{2{\bf k},2{\bf k}'}
=\frac{V_0\left( 4M^2+gg'k_z\, k_z'\right) }
{\left( 4M^2+g^2k_z^2\right) ^{1/2}
\left( 4M^2+g^{\prime 2}k_z^{\prime 2}\right) ^{1/2}}\; ,
\eqno (54)
$$
$$
V_{1{\bf k},2{\bf k}'}=-V_{2{\bf k},1{\bf k}'}
=\frac{2iV_0M\left( gk_z+g'k_z'\right) }
{\left( 4M^2+g^2k_z^2\right) ^{1/2}
\left( 4M^2+g^{\prime 2}k_z^{\prime 2}\right) ^{1/2}}\; ,
\eqno (55)
$$
where $g'=g(k')$.

To find the self energy one needs to calculate the diagrams shown in
Fig.~10. 

\begin{figure} 
\epsfig{file=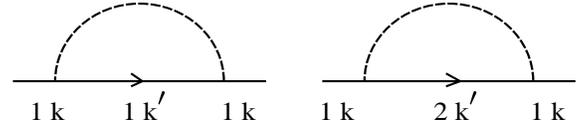,height=4.5cm,width=9cm} 
\caption{Self energy in the case of the Bychkov-Rashba interaction. }
\end{figure}

In what follows we consider the case when SO is small as compared to the
magnetic splitting, so we can expand the self energy
in a small parameter $gk_F/M\ll 1$.
Then, we find
\wide{m}{
$$
\Sigma _1^R(\varepsilon ,{\bf k})\simeq N_i\, V_0^2\left( 1+\frac{g^2k_z^2}{4M^2}\right)
\int \frac{d^3{\bf k}'}{(2\pi )^3}
\left( 1-\frac{g^{\prime 2}k_z^{\prime 2}}{4M^2}\right)
\frac1{\varepsilon -\varepsilon _1({\bf k}')+\mu +i\delta}
+N_i\, V_0^2\int \frac{d^3{\bf k}'}{(2\pi )^3}\, 
\frac{g^{\prime 2}k_z^{\prime 2}}{4M^2}
\frac1{\varepsilon -\varepsilon _2({\bf k}')+\mu +i\delta}\; ,
\eqno (56)
$$
}
and a similar expression for $\Sigma _2^R(\varepsilon, {\bf k})$. 
In these formulas we can put $\varepsilon =0$, $k=k_{F\downarrow (\uparrow )}$, 
and average over the Fermi surfaces. 
This leads to the relaxation times for electrons
corresponding to the bands 1 and 2,
$$
\tau _1=\frac1{2\pi \nu _1\, N_i\, V_0^2}
\left( 1-\frac{\nu _\uparrow }{\nu _\downarrow }
\frac{g_{\uparrow }^2k_{F\uparrow }^2}{12M^2}\right)\; ,
\eqno (57)
$$
$$
\tau _2=\frac1{2\pi \nu _2\, N_i\, V_0^2}
\left( 1-\frac{\nu _\downarrow }{\nu _\uparrow }
\frac{g_{\downarrow }^2k_{F\downarrow }^2}{12M^2}\right)\; ,
\eqno (58)
$$
where $g_{\uparrow }=g(k_{F\uparrow })$ and
$g_{\downarrow }=g(k_{F\downarrow })$.
The above formulae explicitly show
the contribution of SO interaction to the
relaxation time.

We also need to calculate the densities of states $\nu _1$  and $\nu _2$ 
at the Fermi surfaces and to relate them to the ones in the absence of the 
spin-orbit interaction, $\nu _\downarrow $ and $\nu _\uparrow $. 
To do this we write
$$
\nu _{1(2)}=\int \frac{d^3{\bf k}}{(2\pi )^3}\, 
\delta \left( \mu -\varepsilon _{1(2)}({\bf k})\right)\; ,
\eqno (59)
$$
and then, upon integrating over $k_x$, we obtain
$$
\nu _{1(2)}=\frac{(2m)^{1/2}}{(2\pi )^3}\int _{\varepsilon _{1(2)}^l({\bf k})<\mu }
\frac{dk_y\, dk_z}{\left( \mu -\varepsilon _{1(2)}^l({\bf k})\right) ^{1/2}}\; ,
\eqno (60)
$$
where 
$$
\varepsilon _{1(2)}^l({\bf k})=\frac{k_y^2+k_z^2}{2m}\pm s_{\bf k}\; ,
\eqno (61)
$$
$$
s_{\bf k}=\left[ \left( M-gk_y\right) ^2+g^2k_z^2\right] ^{1/2}\; .
\eqno (62)
$$
Expanding in powers of $gk_F/M\ll 1$ and calculating the integrals in (60),
we find
$$
\nu _1=\nu _\downarrow 
\left( 1+\frac{m^2g_{\downarrow }^2}{2k_{F\downarrow }^2}
-\frac{mg_{\downarrow }^2}{2M}\right)\; 
\eqno (63)
$$
and 
$$
\nu _2=\nu _\uparrow 
\left( 1+\frac{m^2g_{\uparrow }^2}{2k_{F\uparrow }^2}
-\frac{mg_{\uparrow }^2}{2M}\right)\; .
\eqno (64)
$$

The relaxation times $\tau _{1}$ and $\tau _{2}$ enter the expressions 
for the renormalized retarded
($G^R$) and advanced ($G^A$) Green functions
$$
G^{R,A}\left( \varepsilon ,{\bf k}\right)
=\frac{\varepsilon -k^2/2m+\mu -M\sigma _z+gk_y\sigma _z-gk_z\sigma _y}
{\left( \varepsilon -\varepsilon _{1{\bf k}}+\mu \pm i/2\tau _1\right)
\left( \varepsilon -\varepsilon _{2{\bf k}}+\mu \pm i/2\tau _2\right) }\; .
\eqno (65)
$$
Now we have already all quantities necessary to
calculate the renormalized vertex
$\Gamma _{\alpha\beta\gamma\delta}$.
The general ladder-type equation for such a vertex reads
$$
\Gamma _{\alpha\beta\gamma\delta}(\omega ,q)
=N_i\, V_0^2\, \delta _{\alpha\delta}\delta _{\beta\gamma}
$$
$$
+N_i\, V_0^2\, \sum_{\nu s}\, \Pi _{\alpha\nu \beta s}(\omega ,q)\,
\Gamma _{\nu s\gamma\delta}(\omega ,q)\; ,
\eqno (66)
$$
where $V_0$ is the matrix element of the short-range impurity potential, 
and
$$
\Pi _{\alpha\nu \beta s}(\omega ,q) 
=\int \frac{d^3{\bf k}}{(2\pi )^3}\;
G_{\alpha\nu}^R\left( \omega ,{\bf k}
+{\bf q}\right) \,   
G_{\beta s}^A\left( 0,\, -{\bf k}\right)\; .
\eqno (67)
$$
As before, we will restrict ourselves by considering the coupling constant 
$g$ to be small as compared to the spin splitting, $gk_F/M\ll 1$.

Using Eqs.~(66) and (67), and integrating over $k_x$, we find for 
$q,\omega =0$      
\wide{m}{
$$
\Pi _{\alpha\nu \beta s}(0,0)
=\frac{i(2m)^{1/2}}{8\pi ^2}
\times \left[ 
\int _{\varepsilon _1^l({\bf k})<\mu }dk_ydk_z
\frac{\left[ s_{\bf k}-M\sigma _z+g(k_y\sigma _z -k_z\sigma _y)\right] _{\alpha\nu}
\left[ s_{\bf k}-M\sigma _z-g(k_y\sigma _z -k_z\sigma _y)\right] _{\beta s}}
{s_{\bf k}(s_{\bf k}+s_{-{\bf k}})(s_{-{\bf k}}-s_{\bf k}+i/\tau _1)
\left[ \mu-(k_y^2+k_z^2)/2m-s_{\bf k}\right] ^{1/2}}
\right.
$$
$$
\left. 
+\int _{\varepsilon _2^l({\bf k})<\mu }dk_ydk_z
\frac{\left[ -s_{\bf k}-M\sigma _z+g(k_y\sigma _z -k_z\sigma _y)\right] _{\alpha\nu}
\left[ -s_{\bf k}-M\sigma _z-g(k_y\sigma _z -k_z\sigma _y)\right] _{\beta s}}
{s_{\bf k}(s_{\bf k}+s_{\bf -k})(s_{\bf k}-s_{\bf -k}+i/\tau _2)
\left[ \mu-(k_y^2+k_z^2)/2m+s_{\bf k}\right] ^{1/2}}
\right] .
\eqno (68)
$$
}
If we put $g=0$ in Eq.~(68), we find that only two matrix elements
are nonzero, namely $\Pi _{\uparrow\uparrow \uparrow\uparrow }(0,0)$ and
$\Pi _{\downarrow\downarrow \downarrow\downarrow }(0,0)$. Assuming 
$Dq^2\tau ,\; \omega \tau \ll 1$, we can just supplement this result by
the terms of the expansion in small $q$ and $\omega $ 
$$
\Pi _{\uparrow\uparrow \uparrow\uparrow }(\omega ,q)
=2\pi \nu _{\uparrow }\tau _\uparrow 
\left( 1-D_\uparrow q^2\tau _\uparrow +i\omega \tau _\uparrow \right) \; ,
\eqno (69)
$$
and a similar  one for 
$\Pi _{\downarrow\downarrow \downarrow\downarrow }(\omega ,q) $.

Let us consider now the limit of weak
SO coupling and expand $\Pi _{\alpha\nu \beta s}(q,\omega )$
in powers of $gk_F/M$. 
The first non-vanishing term of this expansion is quadratic in this parameter.
According to Eq.~(68), the other small parameter in the limit of 
$g\rightarrow 0$ is $gk_F\tau $. 
Thus, if we assume $M\tau \ll 1$, we can neglect the terms
$\sim (gk_F\tau )^2$ and keep the terms $\sim gk_F/M $. 
In the opposite limit, $M\tau \gg 1$ (clean ferromagnet), we should keep
terms $\sim (gk_F\tau )^2$ and neglect terms $\sim gk_F/M $. Since our 
considerations of the localization corrections are limited by  
$\varepsilon _F\tau \gg 1$, the dirty case is possible only for some weak
ferromagnets, when $M\ll \varepsilon _F$ (in other words, for ferromagnets
with very low polarization).

In view of Eq.(49), $gk_F$ is the amplitude of the spin-flip process.
In the classical picture $gk_F$ is the angle of spin rotation
in the unit time. Hence, $gk_F\tau $ is just the angle of the classical 
spin rotation due to the Bychkov-Rashba perturbation at the mean free path 
$l$ of the electron. Thus, the smallness of the parameter $gk_F\tau $ 
corresponds to a small spin rotation angle at the length $l$.       

Since considerations in this paper are restricted to the case
$M\tau \gg 1$, we will not discuss here the opposite limit, $M\tau \ll 1$.
Such a case needs a special treatment.
Indeed, as we pointed out in Sec.~1, only the condition
$M\tau \gg 1$ allows to restrict oneselves to the triplet Cooper channel.
The result on the only nonvanishing matrix elements 
$\Pi _{\uparrow\uparrow\uparrow\uparrow}(0,0)$ and 
$\Pi _{\downarrow\downarrow\downarrow\downarrow}(0,0)$
for $g=0$ (see Eqs. (68) and (69)) also refers
to the case $M\tau \gg 1$ since in Eqs.~(65) and (68) we account for
the self energy as a small shift of poles of the Green function from
the real axis.   

In the limit of small SO interaction, $gk_F\tau \ll 1$, and for
$M\tau \gg 1$, we can neglect in Eq.~(68) all terms of the order of 
$gk_F/M$. As a result, we find only two nonvanishing matrix elements
\wide{m}{
$$
\Pi_{\uparrow\uparrow\uparrow\uparrow}(\omega ,q)\equiv
\Pi _\uparrow (\omega ,q)
=2\pi \nu _{\uparrow }\tau _\uparrow
\left( 1-D_\uparrow \, q^2\, \tau _\uparrow
+i\omega \, \tau _\uparrow
-\frac43 \, g_\uparrow ^2\, \tau _\uparrow ^2\, k_{F\uparrow }^2\right)\; , 
\eqno (70)
$$
$$
\Pi_{\downarrow\downarrow\downarrow\downarrow}(\omega ,q)\equiv 
\Pi _\downarrow (\omega ,q)
=2\pi \nu _{\downarrow }\tau _\downarrow
\left( 1-D_\downarrow \, q^2\, \tau _\downarrow
+i\omega \, \tau _\downarrow
-\frac43 \, g_\downarrow ^2\, \tau _\downarrow ^2\, k_{F\downarrow }^2\right) \; ,
\eqno (71)
$$
which give rise to the Cooperon
$\Gamma _{\uparrow }(\omega ,q)$ in the form of Eq.~(18), with the effective 
spin-orbit relaxation time
$$
\frac1{\tilde{\tau }_{so\uparrow }}
=\frac43 \, g_\uparrow ^2\, k_{F\uparrow }^2\, \tau _\uparrow \; .
\eqno (72)
$$
Similar expressions can be obtained  for
$\Gamma _{\downarrow }(\omega ,q)$ and 
$\tilde{\tau }_{so\downarrow }$.

If we take the limit $gk_F\tau \gg 1$, then for $M\tau \gg 1$ we still can 
consider the SO interaction as a small perturbation, $gk_F/M\ll 1$.
Calculating the integral (68) up to the second order in this parameter,
we find
$$
\Pi _{\alpha\nu\beta s}(\omega ,q)
=\frac{3\pi \nu _{\downarrow }\tau _1}
{8(g_\downarrow k_{F\downarrow }\tau _1)^2}\left[
\left( 1-D_\downarrow q^2\tau _1
+i\omega \tau _1
-\frac{mg_{\downarrow }^2}{2M}
-\frac{g_{\downarrow }^2k_{F\downarrow }^2}{3M^2}
+\frac{m^2g_{\downarrow }^2}{2k_{F\downarrow }^2}\right)
(1-\sigma _z)_{\alpha\nu}(1-\sigma _z)_{\beta s}
\right.
$$
$$
\left.
-\frac{g_{\downarrow }^2k_{F\downarrow }^2}{3M^2}\sigma _{\alpha\nu}^y\sigma _{\beta s}^y
+\frac{g_{\downarrow }^2k_{F\downarrow }^2}{6M^2}\delta _{\alpha\nu}(1-\sigma _z)_{\beta s}
+\frac{g_{\downarrow }^2k_{F\downarrow }^2}{6M^2}(1-\sigma _z)_{\alpha\nu}\delta _{\beta s}
-\frac{mg_{\downarrow }^2}{M}(1-\sigma _z)_{\alpha\nu}(1+\sigma _z)_{\beta s} 
\right]
$$
$$
+\frac{3\pi \nu _{\uparrow }\tau _2}
{8(g_\uparrow k_{F\uparrow }\tau _2)^2}\left[
\left( 1-D_\uparrow q^2\tau _2
+i\omega \tau _2
+\frac{mg_{\uparrow }^2}{2M}
-\frac{g_{\uparrow }^2k_{F\uparrow }^2}{3M^2}
+\frac{m^2g_{\uparrow }^2}{2k_{F\uparrow }^2}\right)
(1+\sigma _z)_{\alpha\nu}(1+\sigma _z)_{\beta s}
\right.
$$
$$
\left.
-\frac{g_{\uparrow }^2k_{F\uparrow }^2}{3M^2}\sigma _{\alpha\nu}^y\sigma _{\beta s}^y
+\frac{g_{\uparrow }^2k_{F\uparrow }^2}{6M^2}\delta _{\alpha\nu}(1+\sigma _z)_{\beta s}
+\frac{g_{\uparrow }^2k_{F\uparrow }^2}{6M^2}(1+\sigma _z)_{\alpha\nu}\delta _{\beta s}
+\frac{mg_{\uparrow }^2}{M}(1+\sigma _z)_{\alpha\nu}(1-\sigma _z)_{\beta s} 
\right]\; .
\eqno (73)
$$
Using (73), we can find all nonvanishing matrix elements of 
$\Pi _{\alpha\nu\beta s}(\omega ,q)$ and then solve the ladder equations of Fig.~7.
We do not make this calculation since we notice that the presence
of the small factor $1/(gk_F\tau )^2$ leads to complete suppresion of the 
Cooperon. Thus, the localization corrections exist only for $gk_F\tau <1$.

Finally, we present the calculation of a classical correction to
the conductivity due to the SO interaction in Hamiltomian (49). 
After calculating the loop diagram with the Green functions (65), and 
using (57),(58),(63),(64), we find
$$
\sigma _{zz}^{(0)}
=\frac{e^2n_\uparrow \tau _\uparrow }{m}\left(
1+\frac{m^2g_\uparrow ^2}{k_{F\uparrow }^2}
-\frac{mg_\uparrow ^2}{M}
-\frac{\nu _\downarrow }{\nu _\uparrow }
\frac{g_\downarrow ^2\, k_{F\downarrow }^2}{12M^2}\right)
+\frac{e^2n_\downarrow \tau _\downarrow }{m}\left(
1+\frac{m^2g_\downarrow ^2}{k_{F\downarrow }^2}
-\frac{mg_\downarrow ^2}{M}
-\frac{\nu _\uparrow }{\nu _\downarrow }
\frac{g_\uparrow ^2\, k_{F\uparrow }^2}{12M^2}\right) \; .
\eqno (74)
$$
}
This expression contains corrections related to renormalization of
the density of states (second and third terms in each bracket) and
of the scattering time (fourth terms). 

We should also take into account
a renormalization of the chemical potential $\mu $. For this purpose
we calculate the total number of spin up and down electrons and 
impose the condition of constant particle number.
After simple calculations we find
the correction to the chemical potential
$$
\Delta \mu 
=\frac{g_\uparrow ^2\, k_{F\uparrow }^3+g_\downarrow ^2\, k_{F\downarrow }^3}
{6M(k_{F\uparrow }+k_{F\downarrow })}
-\frac{m(g_\uparrow ^2\, k_{F\uparrow }+g_\downarrow ^2\, k_{F\downarrow })}
{2(k_{F\uparrow }+k_{F\downarrow })}\; .
\eqno (75)
$$
Calculating the corresponding corrections to Fermi momenta, 
$k_{F\uparrow ,\downarrow }$, 
we find the corrections to spin up and down classical conductivities
connected with renormalization of the chemical potential
$$
\frac{\Delta \sigma _{zz\uparrow }^{(0)}}{\sigma _{zz}^{(0)}}
=\frac{m(g_\uparrow ^2\, k_{F\uparrow }^3+g_\downarrow ^2\, k_{F\downarrow }^3)}
{2Mk_{F\uparrow }^2(k_{F\uparrow }+k_{F\downarrow })}
-\frac{3m^2(g_\uparrow ^2\, k_{F\uparrow }+g_\downarrow ^2\, k_{F\downarrow })}
{2k_{F\uparrow }^2(k_{F\uparrow }+k_{F\downarrow })}\; ,
\eqno (76)
$$
$$
\frac{\Delta \sigma _{zz\downarrow }^{(0)}}{\sigma _{zz}^{(0)}}
=\frac{m(g_\uparrow ^2\, k_{F\uparrow }^3+g_\downarrow ^2\, k_{F\downarrow }^3)}
{2Mk_{F\downarrow }^2(k_{F\uparrow }+k_{F\downarrow })}
-\frac{3m^2(g_\uparrow ^2\, k_{F\uparrow }+g_\downarrow ^2\, k_{F\downarrow })}
{2k_{F\downarrow }^2(k_{F\uparrow }+k_{F\downarrow })}\; .
\eqno (77)
$$
It should be noted that the classical corrections, (76) and (77), can be of any sign
since the inequality $M\gg 1/\tau $ does not necessarily
imply any relations between the magnitude of $M$ and the Fermi energies of 
majority and minority electrons.

The total correction to conductivity includes all corrections to the 
classical part, Eqs.~(74), (76)  and (77), as well as
the quantum correction in
the form of Eqs.~(21) or (36 a,b) with $\tilde{\tau }_{so}$, given
by (72) for $g\, k_F \tau <1$.  
 
\section{Effect of internal magnetic induction in a ferromagnet} 

In our model we have taken into account the effect of magnetic induction ${\bf B}$, 
which enters the kinetic energy of electrons through a vector potential ${\bf A}$
and leads to the suppression of localization corrections, Eqs.~(22) and (36b). 
The total magnetic induction ${\bf B}$ inside a ferromagnet includes the external 
magnetic field ${\bf H}^{ext}$ and the internal magnetic induction ${\bf B}^{int}$,
${\bf B}={\bf H}^{ext}+{\bf B}^{int}$, where 
$B^{int}_i=4\pi \left( \delta _{ij}-n_{ij}\right) M_{0j}$, ${\bf M}_0$ is the 
magnetization vector and $n_{ij}$ is the demagnetizing factor tensor.\cite{ll8}    
The magnitude of the internal magnetic induction may be rather high in strong 
ferromagnets. Nevertheless, weak localization  corrections were observed, e.g., in 
Ni films.\cite{aprili}

We can present some numerical estimations of the magnitude of $B^{int}$. 
For example, in the case of Fe we take\cite{landolt}  $4\pi M_0\simeq 2$~T.
Thus, for the bulk Fe, when the demagnetizing field is negligibly small, we 
have $B^{int}\simeq 2$~T. 
The critical magnetic induction $B^{crit}$, which can totally 
suppress the localizations correction, is determined differently in 3D and 2D cases. 
In the 3D case we can estimated it by\cite{alt82,alt85} $l_B\simeq l$ ($l$ is the 
electron mean free path). For the parameters of very pure bulk Fe
($m\simeq 4\, m_0$ and $\tau \simeq 5\times 10^{-13}$~s),\cite{hood} we
find $l\simeq 4\times 10^{-5}$~cm, and consequently $B_{crit}\simeq 50$~Oe.
This estimation shows that the localization corrections
in pure bulk Fe are totally suppressed by the internal magnetic induction. 
On the other hand, in not so pure metals, or in magnetic alloys, and 
amorphous materials, one can expect a much shorter mean free path, like 
$l\simeq 10^{-6}$~cm,
which is still large enough for the localization corrections to be small ($k_Fl\gg 1$).
In case of $l=10^{-6}$~cm, one finds $B_{crit}\simeq 7$~T, which is significantly
larger than the internal induction of 2~T and, therefore, makes the localization 
corrections observable.

In the case of thin magnetic films, the demagnetizing factor is of crucial importance. 
For example, when the magnetization vector ${\bf M}_0$ is perpendicular to the 
interface, the demagnetizing factor is unity, $n_{zz}=1$ ($z$ axis is perpedicular to
the plane; other components of demagnetizing tensor are very small), and we have
${\bf B}^{int}=0$. 
In this case, the results of Sec.~4 can be applied with the magnetic induction $B$ equal 
to the external magnetic field $H^{ext}$. 
The critical magnetic field, oriented perpendicular to the plane, which suppresses 
the localization correction,  can be estimated as in the 3D case.  

On the other hand, in the case of in-plane magnetization, the demagnetizing factor 
is much smaller than unity, 
and the non-vanishing internal magnetic induction ${\bf B}^{int}=4\pi {\bf M}_0$
is directed along the film.
However, the effect of parallel magnetic induction on the localization corrections
in a strongly two-dimensional system is absent (Sec.~4). 
For a quasi-2D system like a quantum well, the effect of parallel induction is 
non-vanishing but weak. We can write the corresponding expressions for 
localization corrections in the presence of in-plane magnetic induction 
by simply generalizing the results of Refs.~[\onlinecite{alt81,dug}].

If the film thickness $L$ is large with respect to the electron mean free path $l$, 
but still small enough to consider the film as a 2D system,\cite{alt82,alt85}  
$l\ll L\ll \left[ (D\tilde{\tau }_{so})^{-1}+(D\tau _\varphi )^{-1}\right] ^{-1/2}$,
we can obtain, using Ref.~[\onlinecite{alt81}], the dependence of localization 
correction on magnetic induction in the limit of small $B$, i.e., for $l_B\gg L$: 
\wide{m}{
$$
\Delta \sigma (B)-\Delta \sigma (0)
=\frac{e^2}{4\pi ^2}\left[
\ln \left( 1+\frac{L^2D_\uparrow \left( \tilde{\tau }_{so\uparrow }^{-1}
+\tau _{\varphi \uparrow} ^{-1}\right) ^{-1}}{12\, l_B^4}\right) 
+\ln \left( 1+\frac{L^2D_\downarrow \left( \tilde{\tau }_{so\downarrow }^{-1}
+\tau _{\varphi \downarrow} ^{-1}\right) ^{-1}}{12\, l_B^4}\right) 
\right] .
\eqno (78)
$$
It shows a weaker dependence on the magnetic induction, as compare
to the 3D case.
For larger values of $B$, when $l_B<L$, the dependence on magnetic induction
is as for the 3D case, Eq.~(22). The critical value $B_{crit}$ can also be estimated
as for the 3D case. Thus, for thick clean magnetic films, such that $L\gg l$, and 
$l>l_0\equiv [c/(4e\pi M_0)]^{1/2}$, the in-plane internal magnetic induction suppresses the 
localization corrections completely. But in dirty or amorphous thick films with
$l<l_0$, they can be observed. Using the parameters of Fe,
we find   $l_0\simeq 1.8\times 10^{-6}$~cm.

If the film thickness $L$ is smaller than the mean free path $l$ (ballistic regime), we
use the result of Ref.~[\onlinecite{dug}], which can be presented in a simple
form for some intervals of $B$.\cite{beenakker}\\
To avoid combersome formulae, we introduce the following notations
$$
\frac1{\tau _{c\uparrow ,\downarrow }}
=\frac1{\tilde{\tau }_{so\uparrow ,\downarrow }}
+\frac1{\tau _{\varphi \uparrow ,\downarrow}},\; \; \; \; \; \; \; 
L_{c\uparrow ,\downarrow }
=\left( D_{\uparrow ,\downarrow }\tau _{c\uparrow ,\downarrow }\right) ^{1/2 }.
\eqno (79)
$$  
When $L_c,\, l_B^2/L\gg l$, we obtain
$$
\Delta \sigma (B)-\Delta \sigma (0)
=\frac{e^2}{4\pi ^2}\left[
\ln \left( 1+
\frac{l_\uparrow L^3}{16\, l_B^4}\, \frac{\tau _{c\uparrow }}{\tau _\uparrow }\right)
+\ln \left( 1
+\frac{l_\downarrow L^3}{16\, l_B^4}\, \frac{\tau _{c\downarrow }}{\tau _\downarrow }
\right) \right] ,
\eqno (80)
$$
When $L_c,\, l\gg l_B^2/L$,
$$
\Delta \sigma (B)-\Delta \sigma (0)
=\frac{e^2}{4\pi ^2}\left[
\ln \left( 1+
\frac{L^2}{3\, l_B^2}\, \frac{\tau _{c\uparrow }}{\tau _\uparrow }\right)
+\ln \left( 1
+\frac{L^2}{3\, l_B^2}\, \frac{\tau _{c\downarrow }}{\tau _\downarrow }
\right) \right] ,
\eqno (81)
$$
}
and when $l,\, l_B^2/L\gg L_c$,
$$
\Delta \sigma (B)-\Delta \sigma (0)
=\frac{e^2}{4\pi ^2}\, \frac{\pi }{48}\left(
\frac{L_{c\uparrow }L^3}{l_B^4}\, \frac{\tau_{c\uparrow }}{\tau _\uparrow }  
+\frac{L_{c\downarrow }L^3}{l_B^4}\, \frac{\tau_{c\downarrow }}{\tau _\downarrow }  
\right) .
\eqno (82)
$$  
 
We also find that the critical in-plane magnetic induction in the case of $L\ll l$ 
can  be estimated from $l_B\simeq L$, which gives rise to a much larger critical 
value of magnetic induction, $B^{crit}_{thin film}/B^{crit}_{3D}\simeq (l/L)^2\gg 1$. 
In other words, even for clean magnetic films, the in-plane magnetic induction does
not suppress completely the localization corrections, if the film width is sufficiently
small.

In view of the above considerations, the best configuration for observing weak
localization effects in a ferromagnet is to use a thin film with perpendicular easy
axis, and apply a perpendicular magnetic field.

\section{Summary and conclusions}

We have analyzed the localization corrections to electrical 
conductivity in magnetically polarized materials 
with spin-orbit interactions. The strong magnetic polarization excludes
processes with the singlet Cooperon, which are responsible for the 
antilocalization in nonmagnetic materials with SO scattering.
As a result, the quantum correction to conductivity is always
negative in ferromagnets and leads to negative magnetoresistance.

The strength of SO interaction, together with the phase relaxation time 
due to inelastic processes, determine the magnitude of these corrections.
In the case when the SO interaction is associated with scattering
from impurities and/or other defects, the effective SO scattering time $\tilde{\tau }_{so}$
entering the Cooperon, depends on the dimensionality of the system  
and on the magnetization orientation with respect to the plane of the
system  (in the case of two-dimensional
or quasi-two-dimensional systems).
In the case of strongly two-dimensional ferromagnets
with in-plane magnetization, the inverse time 
$1/\tilde{\tau }_{so}$ is zero. This increases the magnitude of 
the localization correction. The vanishing value of of $1/\tilde{\tau }_{so}$ is 
essentially related to the spin-flip scattering of the Cooperon; the 
usual contribution from the spin-conserving scattering is cancelled by 
the spin-flip contribution, which enters the Cooperon with an opposite sign. 
In the quasi-two-dimensional case  both 
contributions are present, but they do not cancel each other.

We have also found the effective spin relaxation time in the case of
Bychkov-Rashba SO interaction. It  contains contributions from
both spin-flip and spin-conserving scattering processes.

We think that good candidates for observations of the localization
corrections are also semiconducting ferromagnets like GaMnAs,\cite{ohno} 
which are recently extensively studied in view of their
possible applications in spintronics.  
Another example is a new  ferromagnetic 
semiconductor CaB$_6$, where a very small magnetization can be combined 
with a small electron density.\cite{young}
 
\section*{Acknowledgements}
One of authors (V.D.) is thankful to J.~Berakdar, A.~Cr\'epieux and E.~Ya.~Sherman 
for stimulating discussions and valuable comments.  
V.D. and J.B. thank the Polish Committee for Scientific Research for a 
support under Grant No.~5~P03B~091~20, and also for a support by NATO
Linkage Grant  No.~977615.

\end{multicols}

\end{document}